\documentclass[twocolumn]{raa}
\usepackage{graphicx,times}
\usepackage{natbib}
\usepackage{amssymb,amsmath}

\bibpunct{(}{)}{;}{a}{}{,}

\def\ms{M_{\odot}}
\def\LX{L_{\rm X}}

\def\M200{M_{200}}

\usepackage[pagebackref=false]{hyperref}

\hypersetup{colorlinks = true, linkcolor = blue, anchorcolor = red, citecolor = blue, filecolor = blue, urlcolor = red}

\begin{document}

\title{Mock X-ray observations of hot gas with L-Galaxies semi-analytic models of galaxy formation}

   \volnopage{Vol.0 (2023) No.0, 000--000}      
   \setcounter{page}{1}          

   \author{Wenxin Zhong 
      \inst{1,2}
   \and Jian Fu \footnote{corresponding author: fujian@shao.ac.cn}
      \inst{1}
   \and Shiyin, Shen 
      \inst{1,3}
   \and Feng, Yuan
      \inst{1,2}
   }

   \institute{Key Laboratory for Research in Galaxies and Cosmology, Shanghai Astronomical Observatory, CAS, 80 Nandan Rd., Shanghai, 200030, China PR; {\it fujian@shao.ac.cn}
        \and
            University of Chinese Academy of Sciences, No. 19A Yuquan Road, 100049, Beijing, China PR
        \and Key Lab for Astrophysics, Shanghai, 200034, Shanghai, China PR}


\abstract{We create mock X-ray observations of hot gas in galaxy clusters with a new extension of L-Galaxies semi-analytic model of galaxy formation, which includes the radial distribution of hot gas in each halo. Based on the model outputs, we first build some mock light cones, then generate mock spectra with SOXS package and derive the mock images in the light cones. 
Using the mock data, we simulate the mock X-ray spectra for \emph{ROSAT} all-sky survey, and compare the mock spectra with the observational results. Then, we consider the design parameters of \emph{HUBS} mission and simulate the observation of the halo hot gas for \emph{HUBS} as an important application of our mock work. We find: (1) Our mock data match the observations by current X-ray telescopes. (2) The survey of hot baryons in resolved clusters by \emph{HUBS} is effective below redshift 0.5, and the observations of the emission lines in point-like sources at $z>0.5$ by \emph{HUBS} help us understand the hot baryons in the early universe. (3) By taking the advantage of the large simulation box and flexibility in semi-analytic models, our mock X-ray observations provide the opportunity to make target selection and observation strategies for forthcoming X-ray facilities.
}

   \authorrunning{Zhong et al. }            
   \titlerunning{Hot gas mock X-ray observations with SAMs}  

   \maketitle
   
\section{Introduction} \label{sec:intro}

According to the $\Lambda$CDM cosmological models and the results from \emph{Planck}, baryonic matter contributes about 4.9\% of the total mass in the universe \citep{Planck2020}, which contains cold baryons locked in galaxies (star, interstellar medium (ISM), black hole etc. \citealt{KravtsovBorgani2012}) and hot baryons in diffuse and ionized phase in circumgalactic medium (CGM) and intracluster medium (ICM). According to observations \citep{Shull2012} and simulation work (e.g \citealt{CenOstriker2006}), the cold baryons contribute less than 15\% of the baryon budgets and hot gas dominates the baryon content in the low-redshift universe. 

The X-ray emitted by hot baryons can test cosmological models and provide important information on the baryon and energy cycles of galaxies and clusters, as well as traces how the dark matter structures assembled in large scale. In the past two decades, a number of surveys by X-ray telescopes \emph{XMM-Newton} and \emph{Chandra} have detected the X-ray emission from hot haloes around galaxies (e.g \citealt{LiWang2013}; \citealt{Li2017}; \citealt{Babyk2018}). \emph{ROSAT} completes the first X-ray imaging all-sky survey in soft X-ray band (RASS, \citealt{Voges1999}) and provide catalogues for thousands of galaxy clusters (e.g \citealt{Piffaretti2011}; \citealt{Finoguenov2020}). The new X-ray telescope, \emph{eROSITA}, have completed the Final Equatorial-Depth Survey (eFEDS) by the end of 2019 \citep{Brunner2022}, which is a verification of the \emph{eROSITA} all-sky survey (eRASS). The catalogue from eFEDS, which includes 542 candidates of galaxy clusters detected as the extended X-ray sources in the 140 $\deg^2$ sky area, helps in the study of CGM and ICM properties (\citealt{Liu2022}; \citealt{Bahar2022}). 

A number of large scale X-ray surveys are proposed to improve our understanding of the hot baryons in the foreseeable future. \emph{eROSITA} will completes eight all-sky surveys in the soft X-ray band by the end of 2023 (eRASS:8), yielding a sample of over $10^5$ galaxy clusters (\citealt{Merloni2012}; \citealt{Predehl2021}). The X-ray survey by \emph{Athena} Phase B will extend the study of hot baryon distributions in ICM by mapping the properties of low-mass groups up to $z\sim2$ (\citealt{Ettori2013}; \citealt{Kaastra2013}). The Wide Field Imager (WFI) survey, during the first four years of operation, is predicted to detect over 10~000 groups and clusters with $z>0.5$, including 20 groups with a mass of $M_{500}\ge5\times10^{13}\ms$ at around $z\sim 2$ \citep{Zhang2020}. The Chinese \emph{HUBS} mission \citep{Cui2020} intends to conduct an all-sky survey of hot baryons in WHIM and CGM with its large field of view and high spectral resolution (see Tab. \ref{tab:hubsparameter} in Sec. \ref{sec:HUBS} of this paper). 

On the other hand, recent cosmological hydrodynamic simulations such as EAGLE (\citealt{Crain2015}; \citealt{Schaye2015}) and Illustris-TNG (\citealt{Springel2018}; \citealt{Nelson2018}) predict hot haloes around groups and clusters. A lot of papers study the X-ray emission from ICM and CGM using the simulation results (e.g \citealt{Stevens2017}; \citealt{Kovacs2019}; \citealt{Martizzi2019}; \citealt{Truong2021}), and some works further make mock X-ray observations based on the plans of X-ray surveys. For example, \citet{Oppenheimer2020} makes predictions of resolved X-ray images for \emph{eROSITA} with EAGLE and Illustris-TNG. \citet{WijersSchaye2022} discusses the detection prospects of X-ray emission lines for \emph{Athena} X-IFU and Lynx Main Array \citep{Gaskin2019} using EAGLE. \citet{Zhang2022} creates mock observations for \emph{HUBS} with Illustris-TNG and assesses the scientific capabilities in detecting extended X-ray emission from hot gas. \citet{Vijayan2022} generates X-ray emission of ISM and CGM from MACER code \citep{Yuan2018}, and simulate \emph{HUBS} observation of elliptical galaxies in four sets of simulations. 

The outputs of semi-analytic models of galaxy formation (hereafter SAMs) offer another choice to build mock observations, such as the mock observatory by \citet{Overzier2013}, mock cones for SKA HI Surveys by \citet{ObreschkowMeyer2014}, and mock galaxy catalogues in multiple bands by \citet{Merson2013}. Due to the low-cost of running SAMs, the main advantage of the SAMs outputs is the large size of the simulation box (e.g the box size of L-Galaxies SAMs is 500~Mpc~$h^{-1}$ based on Millennium Simulation, \citealt{Henriques2015}). The large simulation box helps to construct mock observations in very large sky area without the effect of cosmic variance even at high redshift, i.e the 500~Mpc~$h^{-1}$ box corresponds to over $50\deg^2$ sky area at $z\sim2.0$. The mock catalogue based on Millennium Simulation \citep{Springel2005} can also contain the hot gas sample in very massive haloes ($M_{200}\gtrsim10^{15}\ms$).
On the other hand, the flexibility of SAMs makes it possible to generate multiple mock observations based on outputs with different model parameters and prescriptions, which investigates the effect of physical processes and model parameters on the properties of observational results \citep{SomervilleDave2015}.

In our recent work (\citealt{Zhong2023}, hereafter Paper I), we develop a new extension of L-Galaxies 2015 SAMs \citep{Henriques2015} to study the ionized hot gas in the haloes. In contrast to most previous SAMs work (e.g L-Galaxies 2020 by \citealt{Henriques2020}; DARK SAGE by \citealt{Stevens2016}; Shark by \citealt{Lagos2018}), which mainly focus on the stellar and cold gas components in galaxy disks and ISM, Paper I concentrates on the properties and spatial distribution of hot baryon components, as well as the corresponding X-ray emission from hot gaseous halo. Our model results successfully reproduce various of X-ray observations, like the radial profiles of hot gas temperature, the scaling relations of X-ray luminosity, and the baryon fraction in haloes with different mass.

In this paper, we will create mock X-ray observations of the halo hot gas based on the outputs of the SAMs in Paper I. First, we will build mock light cones using the results of spatial information, and then generate the mock spectra and images in soft X-ray band based on some physical properties. We will consider the device parameters of X-ray facilities to mimic the observations, particularly for the \emph{HUBS} mission. The mock results presented in this paper will aid in target selection and observation strategies optimization for future X-ray surveys of hot gas, and they can also be compared to the mock results from other simulations.

This paper is organized as follows. In Section 2, we will describe the methodology used to create mock X-ray observations for hot gas in the haloes, including the steps to build mock light cones, and the procedures to generate mock spectra and images. We will also show a few examples of the mock images and spectra of galaxy clusters. In Section 3, we will consider the device parameters of X-ray telescopes and simulate the observations based on the mock data. We will simulate mock spectra for \emph{ROSAT} as a benchmark and then focus on the mock observations for \emph{HUBS} mission. In Section 4, we will summarize this paper and look ahead to the future work.


\section{Methods} \label{sec:methods}

In this section, we will describe how to create the mock X-ray observations of hot gas using the model outputs of L-Galaxies SAMs. We will first describe the steps to build the mock light cones, and then the procedures to generate the mock spectra and images of galaxy clusters in the light cones. It should be noted that we do not distinguish between the definitions of galaxy ``group'' and ``cluster'' in the following sections of this paper, which is a collection of galaxies embedded in the same dark matter halo, and we will use the term ``cluster'' for simplicity.

\subsection{Simulation and model samples} \label{sec:samples}

The mock observations in this paper is based on the outputs of the models in Paper I, in which we developed a new branch of the L-Galaxies 2015 \citep{Henriques2015} SAMs to describe the radial distribution of hot ionized gas in ICM and CGM. In Paper I, we use a physical model that takes into account the local instabilities and thermal equilibrium processes for hot gas in the haloes to replace the isothermal sphere in previous models. The model outputs include one-dimensional radial profiles of hot gas density, gas temperature and the bolometric X-ray luminosity profiles around each dark matter halo. The model results successfully reproduce the X-ray observations, such as the radial profiles of hot gas density (e.g the electron density profile from REXCESS by \citealt{Croston2008} and the gas temperature profile from \emph{XMM-Newton} and \emph{Chandra} by \citealt{Bartalucci2017}), scaling relations of X-ray luminosity and temperature (\citealt{Goulding2016} and \citealt{Babyk2018} from \emph{Chandra}, \citealt{Mulchaey2003} and \citealt{Anderson2015} from \emph{ROSAT}, \citealt{Li2016} from \emph{XMM-Newton}), and the baryon fraction in different haloes (\citealt{Gonzalez2013} from \emph{XMM-Newton}, \citealt{Vikhlinin2006} and \citealt{Sun2009} from \emph{Chandra}).

In this paper, the SAMs results used to build the mock observations are based on the dark matter haloes of Millennium Simulation (hereafter MS, \citealt{Springel2005}), which is rescaled to the Planck cosmological parameters ($\Omega_\Lambda=0.685, ~\Omega_m=0.315,~\Omega_{\rm{baryon}}=0.0487, ~\sigma_8=0.829$ and $h=0.673$, \citealt{Planck2020}). The comoving box size of the rescaled MS \citep{Angulo2015} is about 480~Mpc$~h^{-1}$ or 713~Mpc on a side, which is several times larger than recent cosmological hydrodynamical simulations, such as EAGLE (in a box of 100~Mpc) and Illustris-TNG (in a box of 100~Mpc or 300~Mpc). The minimum halo mass is about $2.9\times10^{10}\ms$, which is the mass of 20 simulated particles. The resolution of MS is high enough to mock the observations of hot gas component in most galaxy clusters, and the emission from hot gas in haloes below the resolution is usually undetectable in soft X-ray band. Based on the model results in Paper I, the gas temperature in haloes smaller than $10^{11}\ms$ tends to be lower than 0.1~keV.

The SAMs results are saved as halo and galaxy catalogues in a series of discrete snapshots, each of which corresponds to a certain redshift $z$. Based on the halo merger trees of MS rescaled to the Planck cosmological parameters, the model outputs include 59 snapshots from redshift $z\sim56$ to $z=0$. The catalogues include details of the spatial positions and physical properties of each halo and galaxy.

Based on the model prescriptions in Paper I, the properties of hot gas in each halo, including the gas density $\rho_{\rm hot}$ and bolometric X-ray emission profiles $\LX$ are stored in the form of ``radial profiles'', 
which correspond to the values in a set of spherically symmetrical shells with a certain radius around the halo center. To mimic the real observations, the X-ray luminosity profiles in concentric 3D shells are projected to the surface brightness in 2D rings with
\begin{equation}\label{eq:projection}
I_{X,j} = \frac{1}{A_j}\sum\limits_i {f_{V,ij}} {L_{X,i}},
\end{equation}
in which $L_{X,i}$ (unit: erg~s$^{-1}$) is the bolometric X-ray luminosity in shell $i$, and $I_{X,j}$ (unit: erg~s$^{-1}$~kpc$^{-2}$) is the projected X-ray surface luminosity in ring $j$. $f_{V,ij}$ represents the volume fraction of shell $i$ projected in ring $j$, and $A_j$ is the projected area of ring $j$. The detailed formulae and discussions on the projection can be found in papers like \cite{McLaughlin1999} and \citet{Ettori2002}.

\begin{figure}
\centering
 \includegraphics[angle=0,scale=0.4]{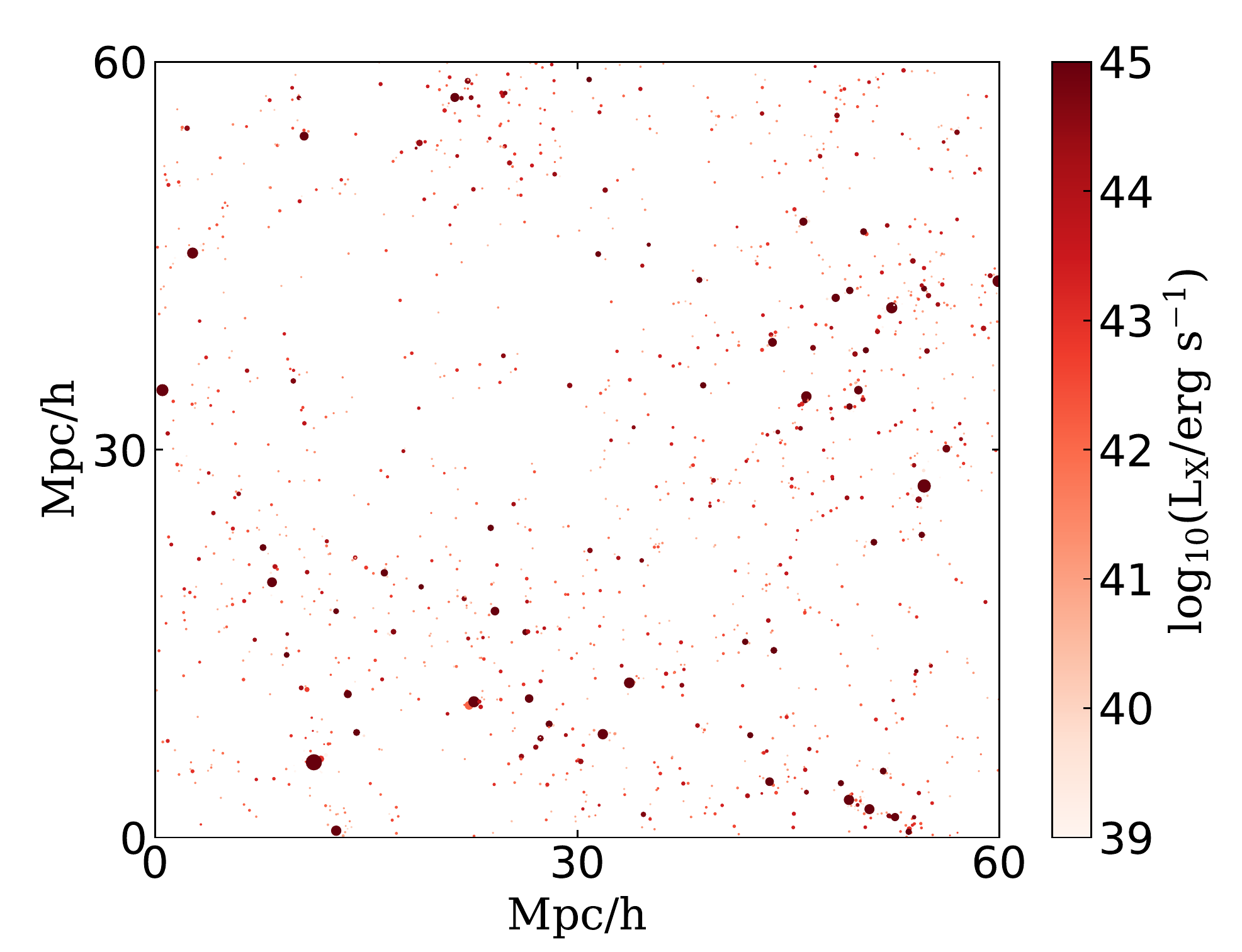}
 \caption{The illustration of hot gas component in the model outputs in a subbox with $1/512$ of the MS volume at $z=0$ (60~Mpc$~h^{-1}$ on a side), in which each dot represents a hot gaseous halo. The size represents the halo radius $R_{200}$ and the color of each dot represents the bolometric X-ray luminosity.
 }\label{fig:simbox}
\end{figure}

In Fig. \ref{fig:simbox}, we show an illustration of hot gas component in the model outputs at $z=0$, which is one of the snapshots used to construct the light cones and mock observations. The illustration is in a subbox of the MS volume with around 60~Mpc$~h^{-1}$ on a side. In this figure, each dot represents one hot gaseous halo, and the size and the color of each dot represents the virial radius $R_{200}$ and the bolometric X-ray luminosity of each halo.

In the framework of SAMs, we have ``halo hot gas'' in the model results and do not distinguish between the ionized hot gas in ICM or CGM, and we only concentrate on the X-ray emission from hot gas components inside the virial radius of each halo in this paper. We should also mention that SAMs do not consider the details of the non-spherical structures such as filaments, knots and cosmic webs. The baryons in these structures are thought to reside in the hot gas halo or the ejecta reservoir out of halo depending on whether they are bounded within the halo potential or not.

\subsection{Light cones} \label{sec:lightcone}

The model results of haloes and galaxies are in cubic simulation boxes at a finite number of redshifts. To mimic the real observation, we convert the cubic boxes into a virtual sky with the spatial information (the 3D positions and 3D velocities). We follow the methods (MoMaF) developed by \citet{Blaizot2005} and \citet{Kitzbichler2007} to create mock catalogues and light cones based on the outputs of SAMs, the details of the methods can be found in the original papers and subsequent works (e.g \citealt{Obreschkow2009}; \citealt{Zoldan2017}). Here, we briefly describe the steps:

\noindent (i) We position the observer at the coordinate origin $(0,0,0)$ and randomly replicate the simulation boxes in a 3D grid. Firstly, we calculate the comoving distance from a box center to the observer and get the corresponding redshift, then we stack the box with the closest redshift.

Due to the relatively large size of the simulation box ($L_{\rm box}\sim710$~Mpc for MS), it is not necessary to use the model outputs in each snapshot at the low redshift. We truncate the 3D grid at $z\sim2$, which includes $8^3=512$ MS boxes. According to the forthcoming plans of X-ray telescopes, $z\sim2$ corresponds to the redshift limit of massive cluster surveys by eRASS \citep{Merloni2012}, and also the redshift limit of the warm-hot baryons and clusters observation by \emph{Athena} \citep{Nandra2013}.

In our current work, we simply splice the boxes at different snapshots together to get a continuous cubic 3D grid and light cones, which is similar to the work by \citet{Zoldan2017} and \citet{Comparat2020}. However, this simplified method may lead to discontinuities in the light cones because of the discrete redshift bins in model outputs. In some mock observation work, the authors interpolate the positions and velocities of the haloes and galaxies between snapshots (e.g \citealt{Merson2013}; \citealt{Smith2022}), and even the intrinsic properties (stellar mass, gas mass, SFR, etc.) of each galaxy \citep{Barrera2022}. According to the results and discussions in \citet{Merson2013} and \citet{Smith2022}, the interpolation mainly affects the results of the galaxy clustering and colour assignment. In this paper, our mock observation mainly focus on the X-ray images and spectra of hot gaseous haloes, and the clustering and distribution in large scale does not affects our mock results. On the other hand, we adopt the energy band with continuous redshift in dealing with the mock images and spectra (see the details in Sec. \ref{sec:spectrum} \& \ref{sec:mockimage}) at high redshift, which avoids producing discrete colour distributions in the results.




\noindent (ii) To suppress spurious radial features caused by the repeated boxes, we assign the ``random tiling'' on the 3D grid, which includes the random operations of shift, rotation and inversion on the 3D coordinates and velocities. 

\noindent (iii) In the stacked 3D grid, we calculate the comoving coordinates $(r_x, r_y, r_z)$ of each object relative to the observer, and convert them to spherical coordinates $(\alpha, \delta, z)$. The right ascension $\alpha$ and declination $\delta$ are calculated by
\begin{equation}\label{eq:radec}
\begin{array}{l}
\alpha  = \arctan \left( {{r_x}/{r_z}} \right)\\
\delta  = \arctan \left( {{r_y}/\sqrt {r_x^2 + r_z^2} } \right).
\end{array}
\end{equation}
Since the mock samples should have a continuous redshift distribution instead of the discrete redshift in the model outputs, we calculate redshift $z$ of each source with its comoving distance ${d_c}=\left(r_x^2 + r_y^2 + r_z^2\right)^{1/2}$ by the equation
\begin{equation}\label{eq:dcredshift}
{d_c}\left( z \right) = \frac{c}{{{H_0}}}\int_0^z {\frac{{dz'}}{{\sqrt {{\Omega _\Lambda } + {\Omega _m}{{\left( {1 + z'} \right)}^3}} }}},
\end{equation}
in which $\Omega _\Lambda$ and $\Omega _m$ are the cosmological parameters. The apparent redshift $z_v$ with Doppler redshift is then calculated by, 
\begin{equation}\label{eq:dopplerredshift}
{z_v} = {z_{\cos}} + \frac{{{v_r}}}{c}\left( {1 + {z_{\cos }}} \right),
\end{equation}
in which $v_r$ is the peculiar velocity projected along line-of-sight, and $z_{\cos}$ is the cosmological redshift in Eq. \ref{eq:dcredshift}.

\noindent (iv) Based on the model outputs in spherical coordinates mentioned above, we create light cones to mimic real observations. Considering the  of \emph{HUBS} (1~$\deg^2$) and \emph{eROSITA} ($1.03^\circ\times1.03^\circ$), we choose ${1^\circ}\times{1^\circ}$ as the angular size of each light cone. The mock data are saved according to the light cones. We generate two sets of light cones: one deep light cone and several shallow light cones. The deep light cone is generated in random direction up to $z\sim2$. The 10 shallow light cones are generated up to $z\sim0.2$\footnote{We choose $z$ up to 0.2 for the shallow light cones, because the comoving distance of $z=0.2$ is just a bit larger than the box size of MS.}, and the center of each shallow light cone is a nearby cluster. Furthermore, It is quite easy to generate more light cones for further statistical analysis.

\begin{figure}
\centering
 \includegraphics[angle=0,scale=0.41]{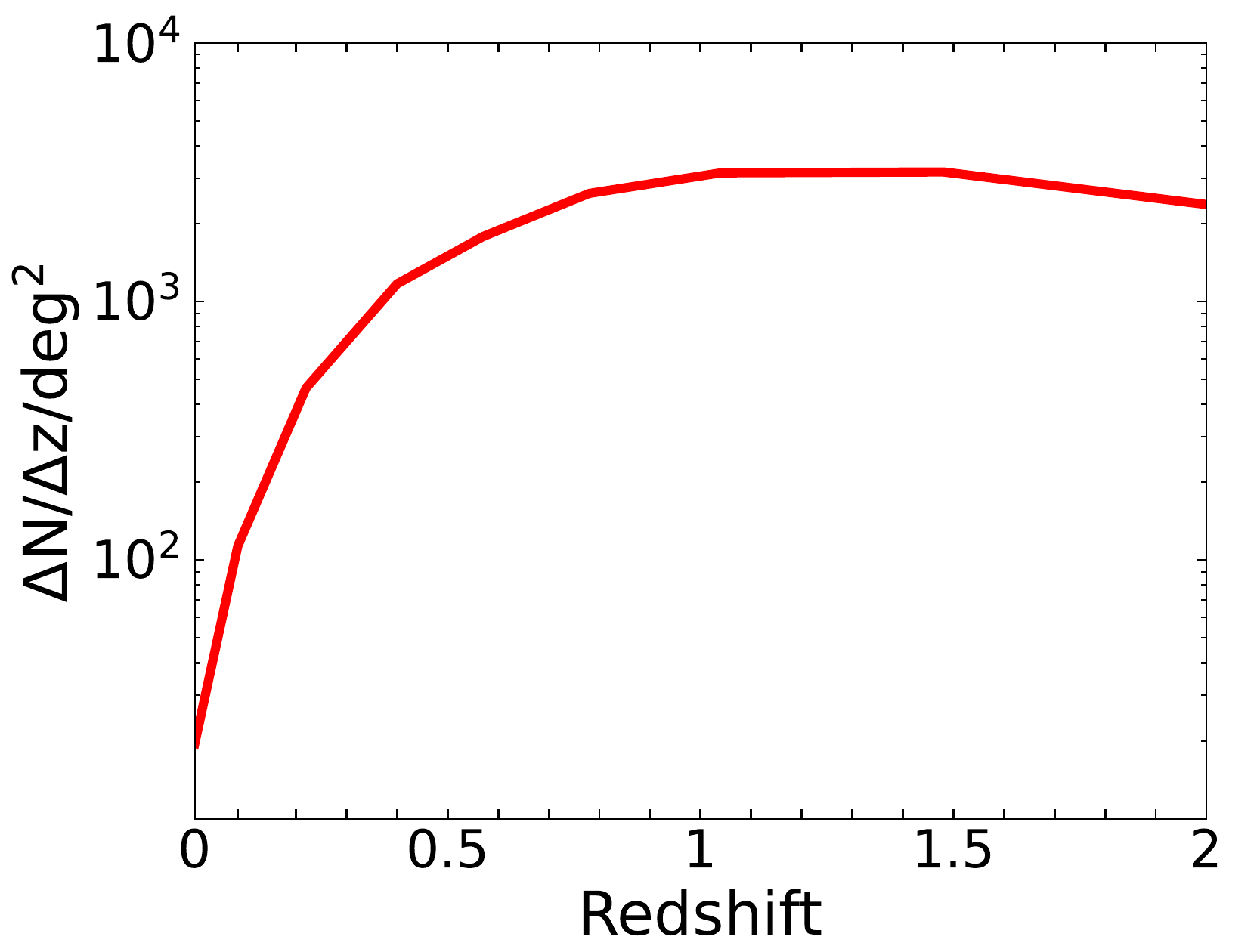}
 \caption{The redshift distribution per redshift bin ($\Delta z=0.2$) per square degree of the haloes with $\M200>10^{12}\ms$, averaged throughout the entire mock sky up to $z=2$. 
 }\label{fig:dndz}
\end{figure}

Based on the model results in Paper I, we focus on the mock data of haloes with $\M200>10^{12}\ms$, since the hot gas temperature in haloes around $10^{12}\ms$ is just above 0.1~keV, and the emission from lower mass haloes is nearly invisible in soft X-ray band. The deep light cone up to $z\sim2$ contains approximately 24~000 haloes above $10^{12}\ms$, and the shallow light cones up to $z\sim0.2$ contains around 73 haloes above $10^{12}\ms$ on average. In Fig. \ref{fig:dndz}, we show redshift distribution per square degree of the haloes with $\M200>10^{12}\ms$ 
, averaged throughout the entire mock sky up to redshift $z=2$. We can see that the halo number per square degree peaks at $z\sim1$ and changes little at higher redshift. 

\subsection{Mock spectra} \label{sec:spectrum}

To generate the mock spectra of the X-ray emission from the halo hot gas, we use the package ``Simulated Observations of X-ray Sources'' (SOXS), whose details can be found in SOXS webpage (\href{sox}{https://hea-www.cfa.harvard.edu/soxs}). In the SOXS package, we apply the APEC spectrum generator based on hot plasmas in collisional ionization equilibrium (CIE) by \citet{Foster2012} and we also consider the Galactic foreground absorption in the spectrum.

In each mock light cone, we generate wide-band spectrum for the hot gas in each halo and also the narrow-band spectra around certain emission lines (e.g the \ion{O}{vii} and \ion{Fe}{xvii} lines). To generate these spectra, the following three properties from L-Galaxies model outputs are used as the input parameters for SOXS: \\
$L_X/4\pi{d_c^2}$: bolometric X-ray flux of hot gas in a halo;\\ 
$T_X$: luminosity-weighted mean gas temperature of a halo; \\
$Z_{\rm gas}$: mean hot gas metallicity of a halo.

The gas metallicity $Z_{\rm gas}$ is defined as the metallicity in hot gas relative to the solar value,
\begin{equation}\label{eq:zgas}
{Z_{{\rm{gas}}}} = \frac{1}{{{Z_ \odot }}}\frac{{{M_{Z,{\rm{hot}}}}}}{{{M_{{\rm{hot}}}}}},
\end{equation}
in which ${{M_{Z,{\rm{hot}}}}}$ is the mass of metal elements in hot phase and ${{M_{{\rm{hot}}}}}$ is the mass of hot gaseous halo. The solar metallicity $Z_{\odot}$ is set to be 0.02. We should note
that the SAMs adopted in this paper does not contain the abundances of different elements but only one value of the total metallicity in hot gas.

For the haloes at high redshift, we make the redshift correction on the mock spectra. Considering $f_o$ and $f_e$ (unit: cnts~s$^{-1}$~keV$^{-1}$~cm$^{-2}$) are the spectra in the observed and emitted-frame, and the relation between $f_o$ and $f_e$ can be written as
\begin{equation}\label{eq:spectrumcorrection}
f_o\left(\nu_o\right)=f_e\left(\nu_o\left(1+z\right)\right),
\end{equation}
in which $\nu_o$ is the frequency in the observed-frame. Then, we get the spectra in the band of the observed-frame.

\subsection{Mock images} \label{sec:mockimage}

To mimic the observations, generating mock images is another important task. Using the projected surface luminosity profile in Eq. \ref{eq:projection} and the mock spectrum in Sec. \ref{sec:spectrum}, we obtain the mock X-ray image for each halo in the light cone.

For a nearby halo with comoving distance $d_c$, the emissivity $S_{\nu}$ (unit: $\rm erg~s^{-1}~cm^{-2}~arcmin^{-1}$) in a given band $\nu$ is
\begin{equation}\label{eq:fluxz0}
S_{\nu} = \frac{A_i}{4\pi d_c^2}\frac{E_{\nu}}{E_{\rm bol}}I_{X,i},
\end{equation}
in which $E_{\nu}$ and $E_{\rm bol}$ (unit: erg~s$^{-1}$) represent the X-ray emission energy in given band $\nu$ and the bolometric energy from the mock spectrum respectively. $I_{X,i}$ (unit: erg~s$^{-1}$~kpc$^{-2}$) from Eq. \ref{eq:projection} is the projected surface brightness of the bolometric luminosity in ring $i$ of the model halo, and $A_i$ is the projected area of ring $i$.

For high redshift haloes in the deep light cone, the redshift correction is made in the calculation of the surface brightness. Similar to the $K$-correction in the magnitude (e.g \citealt{Hogg2002}), the emissivity $S_{\nu_o}$ in a given band $\nu_o$ is
\begin{equation}\label{eq:fluxz}
{S_{{\nu _o}}} = \frac{{{A_i}}}{{4\pi d_c^2}}\frac{{{E_{e,{\nu _o}\left( {1 + z} \right)}}}}{{{E_{e,\rm bol}}}}\frac{{{I_{e,X,i}}}}{{{{\left( {1 + z} \right)}^4}}},
\end{equation}
in which the subscripts $e$ and $o$ represent the quantities in emitted-frame and observed-frame respectively, and the item $(1+z)^{-4}$ represents the redshift correction of the surface brightness. On the other hand, due to the cosmological redshift of the emitter, the observer can detect the X-ray emission from gas with higher temperature in high redshift clusters, i.e
\begin{equation}\label{eq:tgasz}
T_{{\rm gas},e} = \left(1+z\right)T_{{\rm gas},o}.
\end{equation}

With the distribution of $S_{\nu}$, we get the emissivity image for a cluster (see the examples in Sec. \ref{sec:imagespect}). Considering the device parameters of a specific X-ray telescope, such as ARF (ancillary response file), RMF (redistribution matrix file), PSF (point spread function) and exposure time, we can convert the emissivity to photon-count density (in unit: $\rm cnts~arcmin^{-1}$) and generate the mock images for each cluster in the light cones (details can be found in the following sections).

~\\
\begin{figure*}
\centering
 \includegraphics[angle=0,scale=0.8]{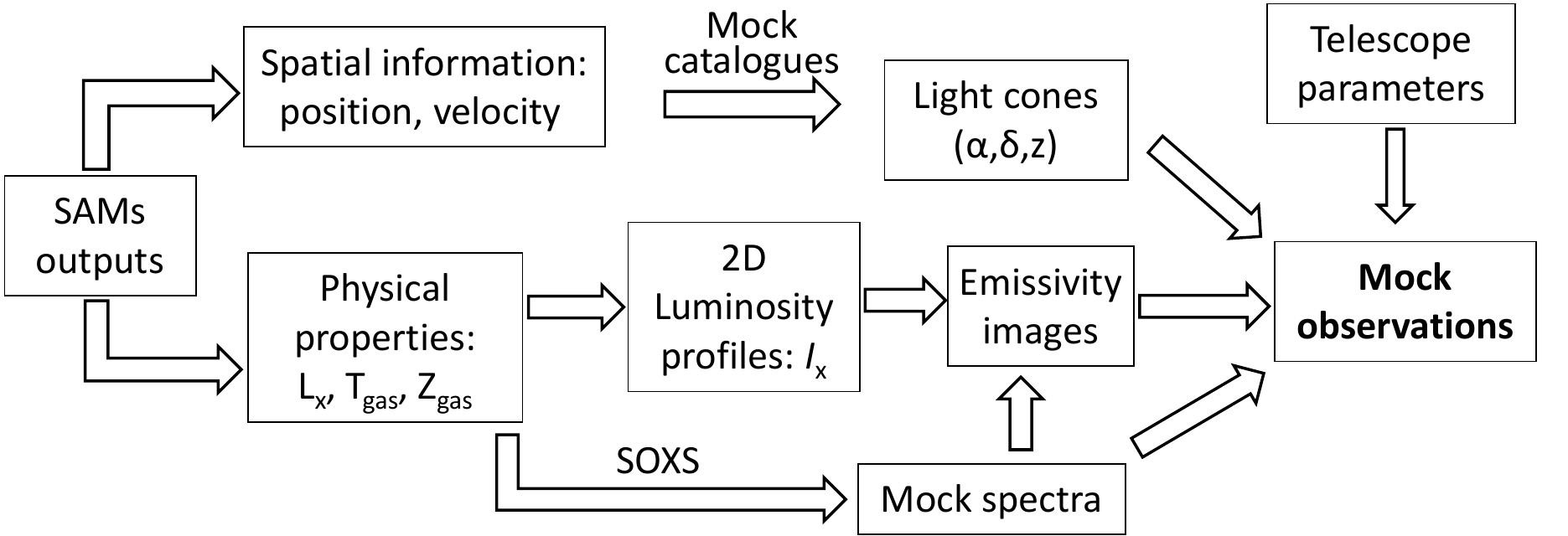}
 \caption{The brief flowchart of the steps involved in creating the mock observations in this paper.
 }\label{fig:flowchart}
\end{figure*}

In summary, we adopt the model outputs from the SAMs in Paper I to create the mock X-ray observations of the hot gaseous haloes. In Fig. \ref{fig:flowchart}, we show a flowchart to describe the steps and procedures in this section. Here we briefly summarize the steps that we follow:

\noindent (i) We adopt the L-Galaxies model outputs running on MS halo merger trees, which are stored in cubic boxes in discrete redshift bins.

\noindent (ii) Based on the spatial information (3D positions and velocities) of each halo, we stack the simulation boxes in 3D grid and assign ``random tiling'' on the grid to suppress the spurious radial features. Then we convert the Cartesian coordinates of each halo to spherical coordinates with respect to the observer.

\noindent (iii) We generate light cones up to different redshift with the angular size of ${1^\circ}\times{1^\circ}$.

\noindent (iv) Using the physical properties (X-ray flux, gas temperature and gas metallicity) from the model outputs, we generate mock X-ray spectra of hot gas in each halo with SOXS packages.

\noindent (v) We project the X-ray luminosity profiles in 3D shells to 2D surface brightness $I_{X,i}$ and derive the X-ray emissivity images with the mock spectra. For the haloes at high redshift, redshift corrections are made on the mock spectra and images.

\noindent (vi) Considering the device parameters, we simulate the observations for X-ray telescopes (see the following sections).

\subsection{Examples of mock images and spectra for clusters} \label{sec:imagespect}

\begin{figure*}
\centering
 \includegraphics[angle=0,scale=0.39]{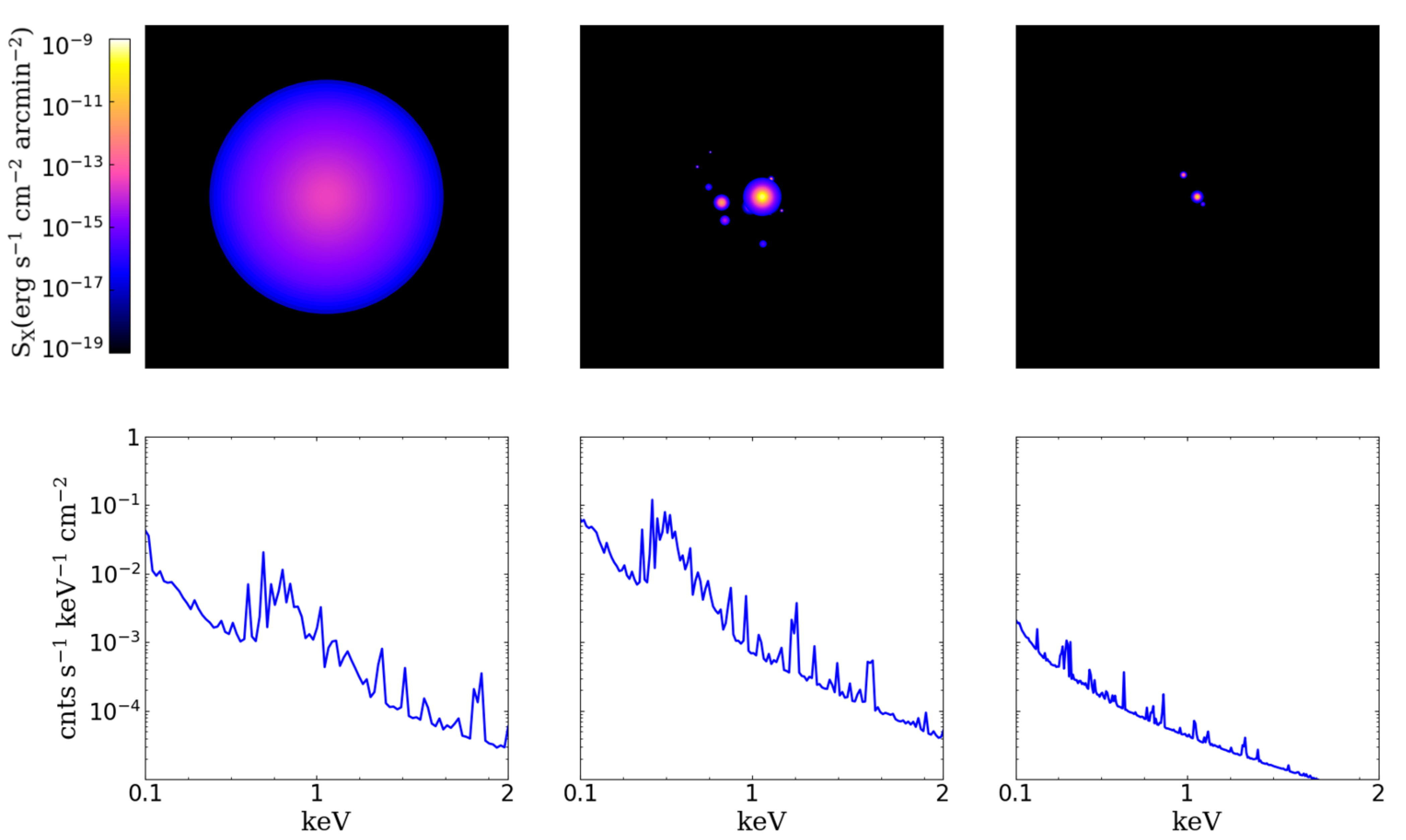}
 \caption{
 Top panels: The X-ray emissivity images in 0.1-2~keV band with ${0.5^\circ}\times{0.5^\circ}$ FoV for model clusters at different redshift. The halo mass and redshift of the 3 clusters are $\M200=10^{12.6}, 10^{14.6}, 10^{14.2}\ms$ and $z=0.03, 0.51, 2.07$ respectively. 
 Bottom panels: The mock spectra of the same clusters in the top panels, generated by SOXS package.
 }\label{fig:imgspec}
\end{figure*}

In this subsection, we will show a few mock images and spectra of hot gas in clusters at different redshift. Considering the methods of contamination removal and member identification for clusters by cross-matching the X-ray sources with samples from multiple-wavelength (e.g \citealt{Salvato2022}), the mock images and spectra shown hereafter are based on the clusters in our mock data. We identify the members of a mock cluster through the halo merger tree in MS, i.e all the central (Type 0\&1 galaxies) and satellite galaxies (Type 2 galaxies) in the subhaloes (the sub-structure within larger virialised halo) of a main FoF halo belongs to one cluster, and the detail definitions of the FoF halo and subhalo can be found in \citet{Springel2005} and \citet{Croton2006}. We define the central galaxy of a cluster as the galaxy in the center of an FoF halo.

\begin{figure}
\centering
 \includegraphics[angle=0,scale=0.35]{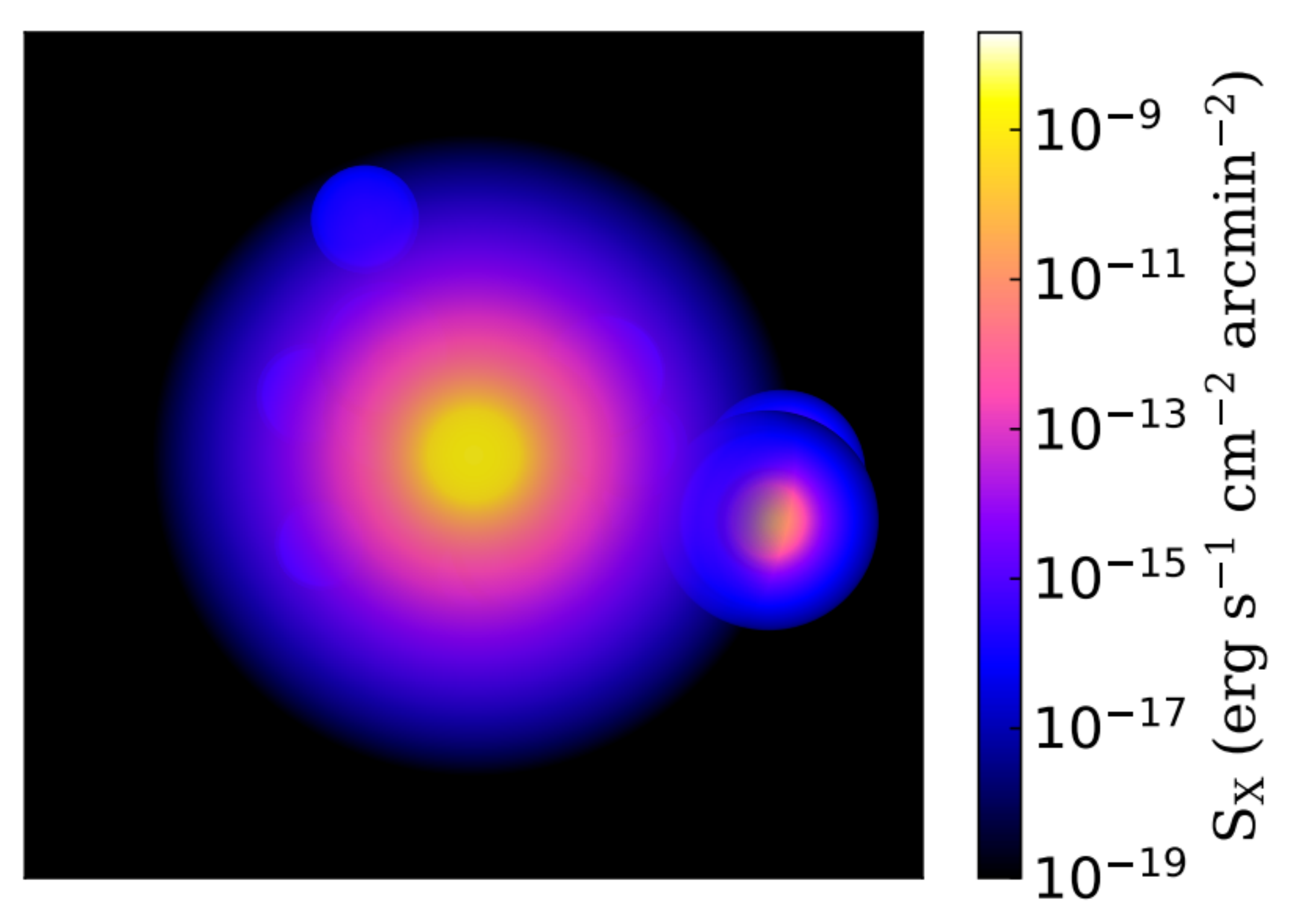}
 \caption{The X-ray emissivity image in 0.1-2~keV band of a nearby rich cluster with $\M200=10^{14.7}\ms$ at $z=0.047$, and the FoV of this image is $1^{\circ}\times1^{\circ}$. The largest source in the center represents the hot gas around the cD galaxy in the main subhalo, and other emission sources represent the hot gas around the satellite galaxies in subhaloes.
 }\label{fig:massiveimg}
\end{figure}

To mimic the observation of the hot gas in nearby and high redshift clusters, we show the examples of mock spectra and emissivity images from three model clusters at different redshift in Fig. \ref{fig:imgspec}. In the left column, we select a cluster with halo mass similar to the Milky Way ($\M200\sim4\times10^{12}\ms$ at $z\sim0.03$) from one of the shallow light cones to mimic the observation of a nearby cluster. For the results at higher redshift, the two mock clusters are from the deep light cone. The middle column is a cluster with $\M200\sim4\times10^{14}\ms$ at $z=0.51$, representing a cluster close to the redshift limit of \emph{HUBS} mission for observation of extended sources (see Sec. \ref{sec:HUBS} for details). In the right column, we select a cluster with $\M200\sim1.6\times10^{14}\ms$ at $z=2.07$, which is around the redshift limit of clusters detection for \emph{eROSITA} and \emph{Athena}. To show the satellite structures more clearly, we also select a cluster with $\M200\sim5\times10^{14}\ms$ at $z\sim0.047$ and show its emissivity image in Fig. \ref{fig:massiveimg}, which represents a nearby rich cluster with a lot of substructures and satellite galaxies around the central galaxy. In each panel of the emissivity images in Fig. \ref{fig:imgspec} \& \ref{fig:massiveimg}, the largest source represents the X-ray emission from the hot gas around the central galaxy and other sources are from the satellite galaxies.

The X-ray emissivity images of these clusters are in 0.1-2~keV band, and the field of view (hereafter FoV) of the images in Fig. \ref{fig:imgspec} is ${0.5^\circ}\times{0.5^\circ}$, while that of the image in Fig. \ref{fig:massiveimg} is $1^{\circ}\times1^{\circ}$. We can see that all the X-ray sources are in spherical shape because the L-Galaxies model assumes a spherically symmetrical profile for each hot gaseous halo. The outer boundary of each emission profile is located at the virial radius $R_{200}$ of each subhalo, and the satellite beyond the outer boundary of the central galaxy belongs to another subhalo. The current version of the L-Galaxies SAMs include the hot baryons beyond the halo potential of a cluster (a.k.a the ejected reservoir). However, the model does not consider the structure and distribution of the unbounded reservoir, so all the mock X-ray emission is from gas inside halo boundary. Although the baryons outside the halo potential are significant, they are very difficult to probe (\citealt{Nicastro2022}, \citealt{Walker2019}), and future model work on the spatial distribution of the unbounded gas in SAMs should be meaningful \citep{Ayromlou2022}.

The emissivity images of nearby clusters in Fig. \ref{fig:imgspec} and \ref{fig:massiveimg} indicates that many X-ray facilities are capable of detecting the structures like the spatial distribution of satellites, and the spatially-resolved spectrum of the entire cluster.
For the cluster at $z\sim2$ in the right column of Fig. \ref{fig:imgspec}, the angular size is around 2~arcmin, which is around the limit of \emph{HUBS} ($1.0\arcmin$ angular resolution, \citealt{Cui2020}), while \emph{eROSITA} ($15\arcsec$ angular resolution, \citealt{Merloni2012}) and \emph{Athena} ($5\arcsec$ angular resolution, \citealt{Kaastra2013}) have the ability to resolve the hot gas in the central and large satellite galaxies.

The bottom three panels of Fig. \ref{fig:imgspec} present the mock X-ray spectra of the same clusters shown in the top panels. In each panel, we stack all the spectra from the central and satellite galaxies together to get a single spectrum for each cluster. In the left panel of the nearby cluster, we can see bumps in 0.5-1.0~keV band in the spectrum, which are the emission lines of the elements O, Fe, Ne, Mg etc, and we will show the detail of these emission lines in the narrow-band spectra in Sec. \ref{sec:HUBS}. In the right panel, 
the relative high gas temperature (mean $T_{\rm gas}\sim2.5$~keV) in this massive halo leads to high ionization fraction for some elements and weak plasma emission lines in the spectrum. 

On the other hand, the high-redshift clusters extend the spectrum of the emitted-frame to the band of high-energy processes, like the AGN and black hole accretion. The current L-Galaxies SAMs do include the prescriptions of gas accretion and AGN feedback processes by central black holes (a.k.a the radio-mode accretion), but the X-ray emission from AGN and black holes is not included. Some works suggest that AGN feedback affects the X-ray luminosity of haloes to some extent. \citet{Gaspari2014} shows that the action of purely AGN feedback is to lower the luminosity and heat the gas. \citet{Puchwein2008} obtains that AGN feedback significantly reduces the X-ray luminosities of poor clusters and groups.
Thus, to get more accurate mock X-ray observations for high redshift clusters, it is important to do future work on the prescriptions of the X-ray emission from black hole accretion and AGN feedback in SAMs.

\section{Mock observations for X-ray telescopes}

In this section, we will consider the device parameters of real X-ray facilities and simulate the observations of hot gas based on our mock data. As a benchmark, we will first simulate the X-ray spectra of \emph{ROSAT} all-sky survey and compare the mock results with the observations. Then, we will focus on the mock observations for the future \emph{HUBS} mission.


\subsection{Mock spectra of clusters for all-sky survey} \label{sec:neprofile}

In this subsection, we will simulate the spectra of the clusters in the first X-ray all-sky survey (RASS) by \emph{ROSAT} \citep{Voges1999} as an application of our mock spectra.

Following the procedures in \citet{Dai2007}, we select clusters from the mock sky up to $z\sim0.2$ and put them to a common distance of 100~Mpc to normalize the apparent luminosity. In \citet{Dai2007}, the clusters of RASS are divided into several groups according to the optical richness, and the richness parameter $N_{*666}$ has a fitting relation with the bolometric X-ray luminosity $L_X$ of a cluster
\begin{multline} \label{eq:n666}
{N_{*666}} = {10^{0.43 \pm 0.03}}{\left( {\frac{{{L_X}}}{{{{10}^{43}}{\rm{erg~s^{-1}}}{h^{-2}}}}} \right)^{0.63 \pm 0.04}}.
\end{multline}
Similarly, our mock clusters are divided into four groups based on $L_{X}$, and the parameters of each group are listed in Tab. \ref{tab:clusters}

\begin{table}
    \centering
    \begin{tabular}{c|c|c|c}
    \hline\hline
      Group & Richness & X-ray luminosity  & Galactic absorption \\
            & $N_{*666}$ &($L_X/10^{42}$~erg~s$^{-1}$) & ($N_{\rm H}/10^{20}$cm$^{-2}$) \\
      \hline
      1 &0.3-1.0 & 0.32-0.97  & 0.7 \\
      2 &1.0-3.0 & 0.97-5.57  & 1.4 \\
      3 &3.0-10 & 5.57-37.7  & 1.8 \\
      4 &10-50 & 37.7-484  & 2.8 \\
    \hline\hline
    \end{tabular}
    \caption{The richness parameter mentioned in \citet{Dai2007}, the corresponding bolometric X-ray luminosity and the Galactic foreground absorption column density for each group of the clusters in Fig. \ref{fig:clusterspectra}.}\label{tab:clusters}
\end{table}

Since the RASS images have already corrected the exposure times for the effects of vignetting, we use the on-axis effective area $A_{\rm eff}$ from Table 5.3 in the \emph{ROSAT} handbook\footnote{https://heasarc.gsfc.nasa.gov/docs/rosat/ruh/handbook/node122.html} to generate the mock spectra comparable with RASS results, then the power $F(\nu)$ (unit: cnts~s$^{-1}$~keV$^{-1}$) received by \emph{ROSAT} at frequency $\nu$ is
\begin{equation} \label{eq:spectrumrosat}
F(\nu)= f(\nu)*A_{\rm eff}(\nu),
\end{equation}
in which $f(\nu)$ (unit: cnts~s$^{-1}$~cm$^{-2}$~keV$^{-1}$) is the flux of a mock spectrum generated by SOXS package. In addition, the Galactic foreground absorption is considered when we calculate the $f(\nu)$ in Eq. \ref{eq:spectrumrosat}, and the column densities $N_{\rm H}$ of the foreground absorption are listed in the right column of Tab. \ref{tab:clusters}, which are same as the values used in \citet{Dai2007}.

\begin{figure*}
\centering
 \includegraphics[angle=0,scale=0.5]{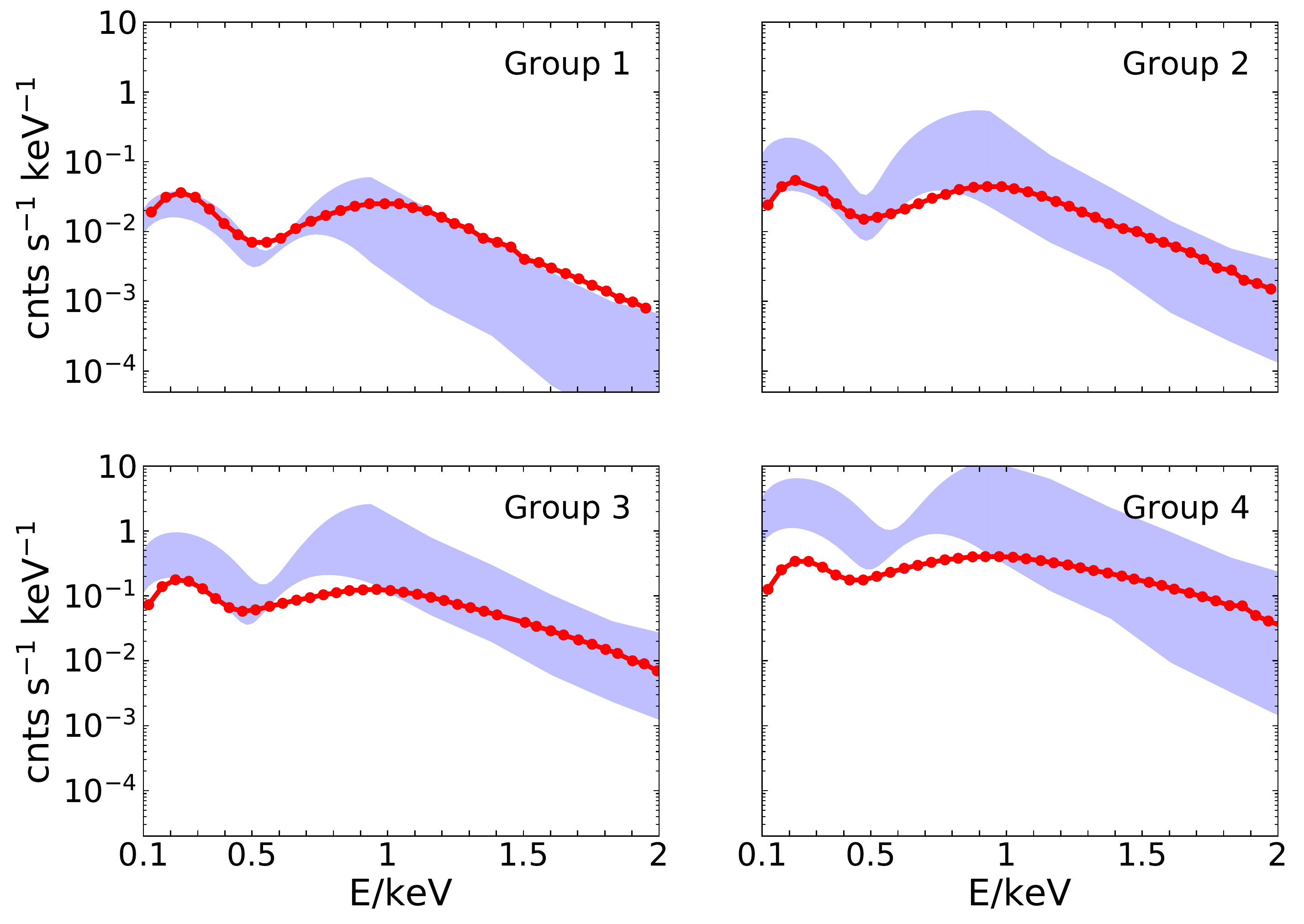}
 \caption{The comparison of the X-ray spectra from the mock clusters and the results from RASS. The mock and observational samples are divided into four panels according to the values of $L_X$ and $N_{*666}$ in Tab. \ref{tab:clusters} respectively. In each panel, the shaded area is the mock spectrum within $\pm1\sigma$ deviations around the mean values for the mock samples, and the red curve is the average spectrum stacked from RASS data by \citet{Dai2007}.
 }\label{fig:clusterspectra}
\end{figure*}

Fig. \ref{fig:clusterspectra} shows the X-ray spectra in 0.1-2~keV band derived from the mock clusters together with the observational spectra from RASS by \citet{Dai2007}, and the samples are divided into four groups according to $L_X$ and $N_{*666}$ in Tab. \ref{tab:clusters}. In each panel, the troughs in the spectra at around 0.5~keV is caused by the drop in the sensitivity of \emph{ROSAT} between 0.3 and 0.6~keV, and the drop at $E\lesssim 0.2$~keV is caused by the foreground absorption.

As shown in Fig. \ref{fig:clusterspectra}, the mock spectra can roughly match the results from RASS in 0.1-2~keV band, and the main difference exists in Group 4 (clusters with $L_X\gtrsim10^{43.5}~\rm erg~s^{-1}$). In these bright clusters, the flux of the mock sample at $E<0.5$~keV is slightly higher than that of RASS, which means the model predicts lower gas temperature $T_X$ than observations in massive haloes. For the clusters in Group 4, the average gas temperature from RASS cluster is $T_X=4.7_{-0.7}^{+1.4}$~keV, while $T_X=2.7$~keV for the mock sample. According to the scaling relations of the hot gas in Paper I (Detailed discussions on the scaling relations of hot gas can be found in Section 3.2 of Paper I.), the inconsistency in the bright clusters is primarily caused by the too steep slope of $L_X-T_X$ relation in the mock clusters. In order to fit the relation ${L_X}\propto{T_X}^{4.5\pm0.2}$ for early type galaxies in the range $L_X\sim 10^{38}\text{-}10^{43}~\rm erg~s^{-1}$ from \emph{Chandra} by \citet{Babyk2018}, the model result in Paper I  gives $L_{X,\rm bol}\sim T_X^{4.5}$, which is steeper than the slopes of the clusters in RASS (${L_X}\propto{T_X}^{2.7\pm0.7}$, \citealt{Dai2007}) and eFEDS ($L_{X,\rm bol}\propto{T_X}^{3.01}$, \citealt{Bahar2022}). On the other hand, according to the discussion in the end of Sec. \ref{sec:imagespect}, the AGN feedback suppresses the X-ray luminosity to some extent. Since the current model does not contain the X-ray emission from AGN, it should be another cause of the discrepancy of the $L_X-T_X$ relation in massive clusters. Future work is necessary to improve the model prescriptions in the bright clusters with $L_X\gtrsim10^{43}$~erg~s$^{-1}$.

\subsection{Mock observations for \emph{HUBS}} \label{sec:HUBS}

\emph{HUBS} (The Hot Universe Baryon Surveyor) is a mission scheduled to launch around 2030 in China. 
Thanks to its large $1\deg^2$ FoV, \emph{HUBS} is at least an order of magnitude more capable of detecting diffuse emission from hot gas than small-FoV X-ray telescopes, which is thought to hide in CGM and IGM. According to the observing strategy \citep{Cui2020}, \emph{HUBS} plans to observe nearby galaxies and clusters with quite long exposure time ($\sim 1$~Ms), and the selection of targets is a very important task to achieve the science objects.
On the other hand, the main advantages of SAMs are the large simulation box and the flexibility to investigate the effect of physical processes. In this section, we will simulate the images and the spectra observations of \emph{HUBS} using our mock data of hot gas based on SAMs. This is an important application of our mock work, which may aid in optimizing future observations for \emph{HUBS} mission.


\begin{table}
    \centering
    \begin{tabular}{c|c}
    \hline\hline
      Parameter & Value \\
      \hline
      Effective area (cm$^2$) & 0-500 \\
      \hline
      Field of view ($\deg^2$) & 1.0 \\
      \hline
      Spectral band (keV)     & 0.1-2.0\\
      \hline
      Energy resolution (eV)  & \\
      Regular                 & 2.0 \\
      Central                 & 0.6 \\
      \hline
      Angular resolution
      (arcmin)                & 1.0 \\
    \hline\hline
    \end{tabular}
    \caption{The key design parameters of \emph{HUBS} mission used to simulate the observations in Sec. \ref{sec:HUBS}.}\label{tab:hubsparameter}
\end{table}

To simulate the observations of \emph{HUBS} mission, we adopt the key design parameters in \citet{Cui2020}, which are shown in Tab. \ref{tab:hubsparameter}. It should be noted that the effective area $A_{\rm eff}$ in Tab. \ref{tab:hubsparameter} is a function of energy from the ancillary response file (ARF) by
\citet{Zhang2022}\footnote{We obtain the ARF file from Zhang et al. through private communication, and the current mock work in Sec. \ref{sec:HUBS} does not include the effect of the RMF (redistribution matrix file).}, which is used to convolve with the flux $f(\nu)$ to get the photon counts. In each light cone, we generate the wide-band and narrow-band mock spectra for each cluster, in which the wide-band spectra are in 0.1-2~keV band with a regular energy resolution of 2~eV while the narrow-band spectra are in the bands around emission lines with a resolution of 0.6~eV. Considering the effective area, angular resolution, and exposure time of \emph{HUBS}, we derive the photon-count images in both wide and narrow bands using the emissivity map and the spectra of each cluster. 

In the generation of mock images and spectra, the foreground and background are also included. According to the results from \emph{XMM-Newton} by \citet{Lumb2002}, the cosmic unresolved X-ray background (hereafter XRB) emission is modeled with a power-law spectrum
\begin{equation}\label{eq:background}
{S_{\rm{b}}} = \left( {9.03 \pm 0.24} \right){\left( {\frac{E}{{{\rm{keV}}}}} \right)^{ - 1.42 \pm 0.03}},
\end{equation}
in which the XRB flux density $S_{\rm b}$ is in the unit of $\rm{cnts~cm^{-2}~s^{-1}~keV^{-1}~sr^{-1}}$. Considering the effective area $A_{\rm eff}$ of \emph{HUBS}, the background count rate can be calculated by
\begin{equation}\label{eq:backgroundrate}
{n_{\rm{b}}} = \int {{S_{\rm{b}}}\left( E \right){A_{{\rm{eff}}}}\left( E \right)dE}
\end{equation}
in which $A_{\rm eff}$ is a function of energy from ARF by \citet{Zhang2022}. Then, we get the value of $n_{\rm{b}}$ in 0.1-2.0~keV band
\begin{equation}\label{eq:nbhubs}
n_{\rm b}=6.3\times10^{-4}\rm{~cnts~arcmin^{-2}~s^{-1}}. 
\end{equation}

For the foreground, we assume a constant column density $N_{\rm H} = 2\times{10^{20}}\rm{cm}^{-2}$ \citep{Willingale2013} for the Galactic foreground absorption, which mainly affects the band below 0.3~keV. 

\begin{figure*}
\centering
 \includegraphics[angle=0,scale=0.45]{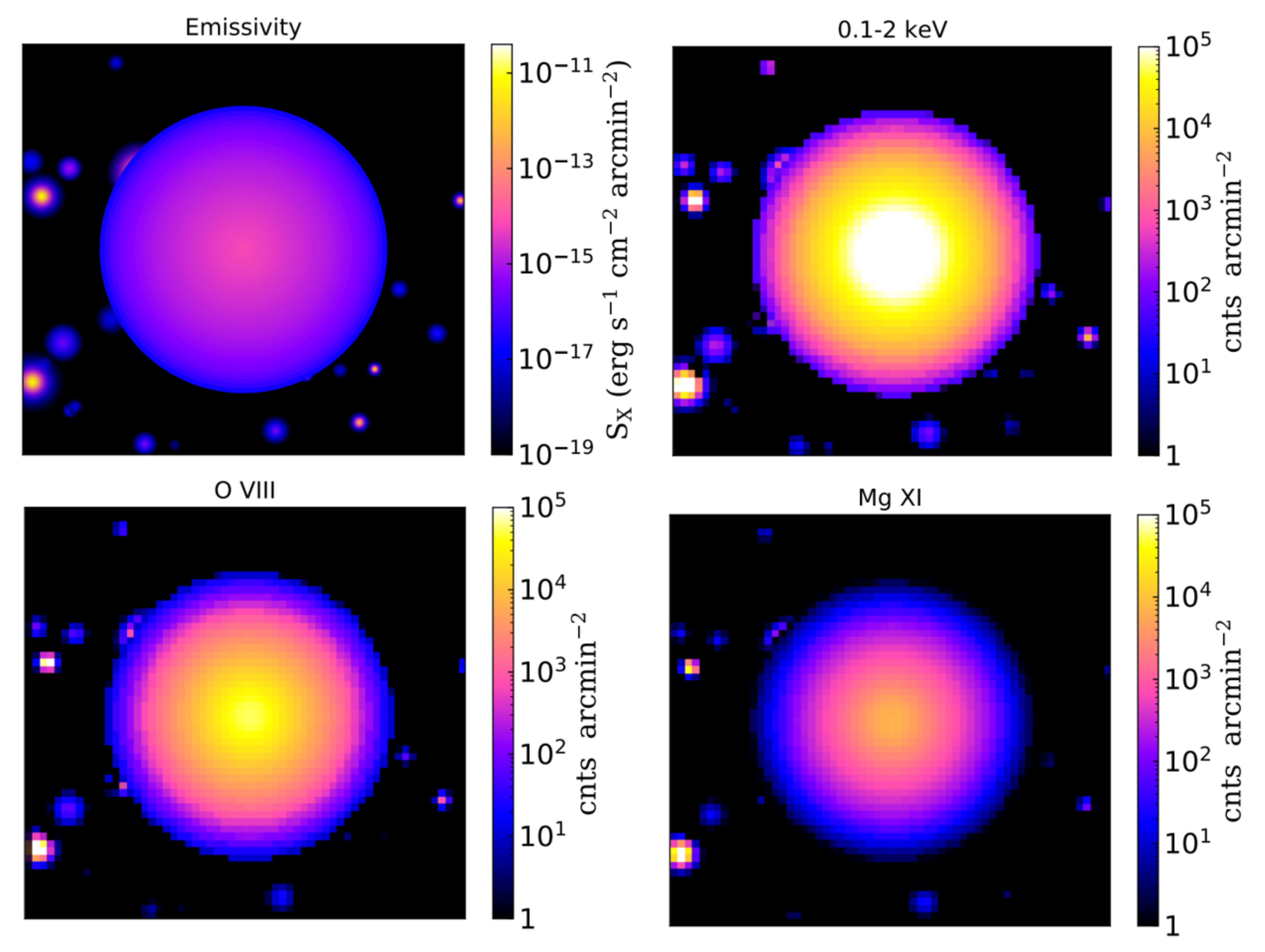}
 \caption{The X-ray mock images of a light cone up to redshift 0.2 with $1^{\circ}\times1^{\circ}$ FoV , and the center of the light cone is a cluster with $\M200\sim5\times10^{12}\ms$ at $z\sim0.014$. The top left panel is an emissivity image in 0.1-2~keV band. The top right panel is a mock image for \emph{HUBS} in 0.1-2~keV band with an exposure time $10^6$~s. The bottom two panels are the corresponding narrow-band images around the \ion{O}{viii} and \ion{Mg}{xi} emission lines with 5~eV bandwidth. The photons from XRB are not included in these images to improve the contrast.
 }\label{fig:hubsimg}
\end{figure*}

In Paper I, our model results predict that the haloes between $10^{12}$ and $10^{13}\ms$ tend to contain a large fraction of hot gas with temperature below 0.5~keV, which is hard-to-detect by many X-ray facilities \citep{Paerels2008}. Thus, we select a nearby cluster with $\M200\sim5\times10^{12}\ms$ at $z\sim0.014$ from one of the shallow light cones, and show the X-ray mock images in Fig. \ref{fig:hubsimg}. This is a typical cluster in our mock sample, representing a local group sized halo in the nearby universe, which is similar to a potential target of \emph{HUBS} mission \citep{Cui2020}. 

The four panels of Fig. \ref{fig:hubsimg} show the emissivity image, wide-band image, and narrow-band images around the \ion{O}{viii} and \ion{Mg}{xi} emission lines respectively. The emissivity and wide-band images are in 0.1-2~keV band, and the wide-band and narrow-band images are calculated with an exposure time of $10^6$~s. In addition to the cluster in the center of the light cone, the images also include the emission from other sources in the $1\deg^2$ FoV. We should note that all the sources in the mock images only contain the X-ray photons from hot gas, and our current SAMs outputs do not contain X-ray emission from other sources, such as AGN and X-ray binaries.

To improve the contrast, the photons from XRB are not shown in Fig. \ref{fig:hubsimg}. Considering the background count rate of \emph{HUBS} in Eq. \ref{eq:nbhubs}, the images in Fig. \ref{fig:hubsimg} show that \emph{HUBS} is capable of detecting the X-ray emission from most of the hot gas inside the virial radius $R_{200}$ of the nearby cluster with an exposure time of $10^6$~s.

\begin{figure*}
\centering
 \includegraphics[angle=0,scale=0.355]{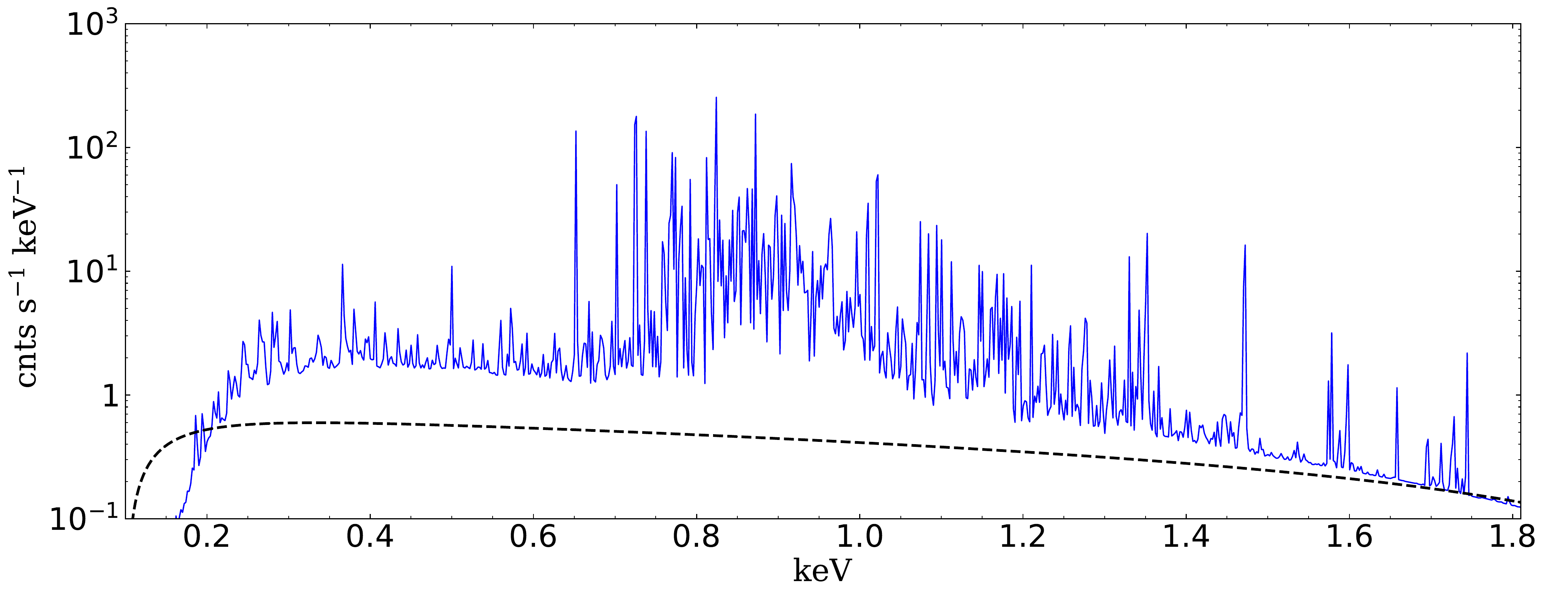}\\
 \includegraphics[angle=0,scale=0.3]{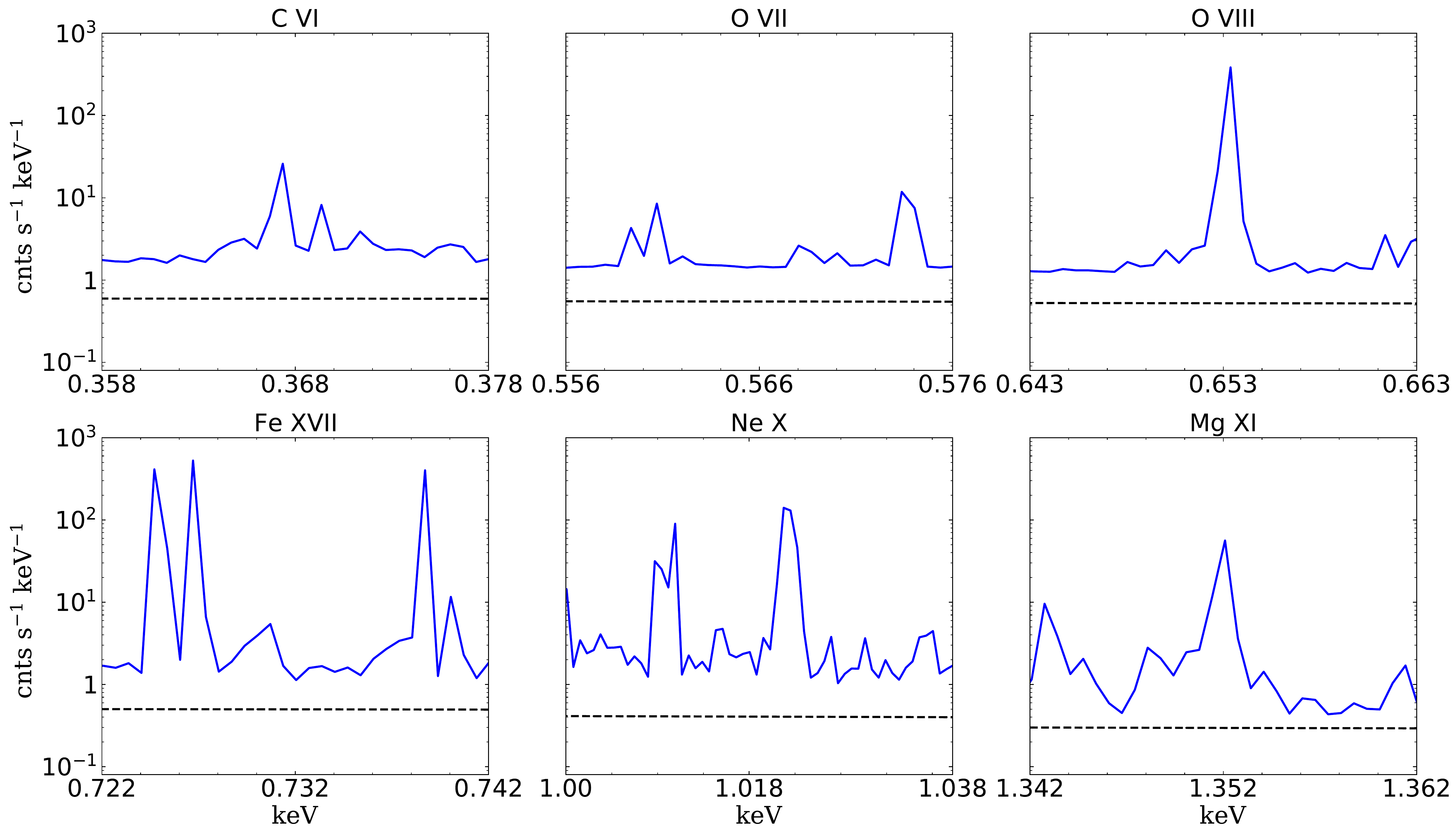}
 \caption{The X-ray mock spectra of the cluster in the center of each panel in Fig. \ref{fig:hubsimg}. The top panel is the spectrum in 0.1-2~keV band with a spectrum resolution of 2~eV, and the bottom panels are the narrow-band spectra around \ion{C}{vi}, \ion{O}{vii}, \ion{O}{viii}, \ion{Fe}{xvii}, \ion{Ne}{x} and \ion{Mg}{xi} emission lines with a spectrum resolution of 0.6~eV. In each panel, the dashed curve shows the spectrum of the photons from XRB. The drop in the left end of the wide-band spectrum is caused by the Galactic foreground absorption and the decrease of effective area below 0.3~keV band.
 }\label{fig:hubsspec}
\end{figure*}

Fig. \ref{fig:hubsspec} shows the X-ray spectra of the cluster in the center of Fig. \ref{fig:hubsimg}. To mimic the contamination sources in the spectra, we include the contributions of all the sources in the solid angle of $R_{200}$ to the cluster center in the light cone, and superpose the redshifted spectra from the contamination sources to the spectrum of the central cluster. The top panel shows the wide-band spectrum in 0.1-2~keV and the bottom panels show the zoomed-in narrow-band spectra around the emission lines of \ion{C}{vi}, \ion{O}{vii}, \ion{O}{viii}, \ion{Fe}{xvii}, \ion{Ne}{x} and \ion{Mg}{xi}. In each panel, the dashed curve shows the spectrum of the XRB, which represents the background noise in the spectrum.

With the help of the central array with high spectral resolution (0.6~eV resolution for the $12\times12$ small-pixel subarray in the center), Fig. \ref{fig:hubsspec} indicates that \emph{HUBS} has the ability to resolve the typical emission lines in nearby clusters, which can be used to study the properties of hot gas, such as the gas temperature, chemical abundance, as well as to trace the baryon cycles in the cluster environment.

Based on the mock data, we can also predict the number of the sources detectable by \emph{HUBS} at different redshift. We assume that most baryons in a cluster can be detected if the signal-to-noise ratio ($\rm S/N$) is greater than 10 inside the radius $R_{500}$ of the halo ($R_{500}$ is the radius within which the density of a halo is 500 times the comic critical density at the halo's redshift). Assuming $n_{\rm s}$ and $n_{\rm b}$ are the count rates of source and background, the signal-to-noise ratio can be calculated by
\begin{equation}\label{eq:snratio}
\rm{S/N} = \frac{n_{\rm s}}{\sqrt {n_{\rm s}+n_{\rm b}}}.
\end{equation}
Considering the criterion $\rm S/N>10$ and the background count rate of \emph{HUBS} in Eq. \ref{eq:nbhubs}, most of the hot baryons of a cluster can be detected if the source count rate in 0.1-2~keV band at $R_{500}$ meets
\begin{equation}\label{eq:countrate}
n_{\rm s}>3\times10^{-4}\rm{~cnts~arcmin^{-2}~s^{-1}}.
\end{equation}

On the other hand, before the PSF (point spread function) of \emph{HUBS} is finally determined, we assume that a cluster can be resolved as an extended source if it exhibits variation in the radial profile of the X-ray luminosity. Based on the $1~\rm arcmin^{2}$ pixel size of \emph{HUBS} and the gas density profiles in the model results (see the results in Section 2.1 of Paper I and also in the paper \citealt{Sharma2012}), we assume a cluster to be an extended source if the angular diameter of its $R_{500}$ is greater than $3~\rm arcmin$. 
For clusters with smaller angular size but $\rm S/N>10$ inside $R_{500}$, they are point-like sources that \emph{HUBS} cannot resolve, but it is still possible for \emph{HUBS} to detect the hot baryons in these clusters with long enough exposure time.


\begin{figure}
\centering
 \includegraphics[angle=0,scale=0.45]{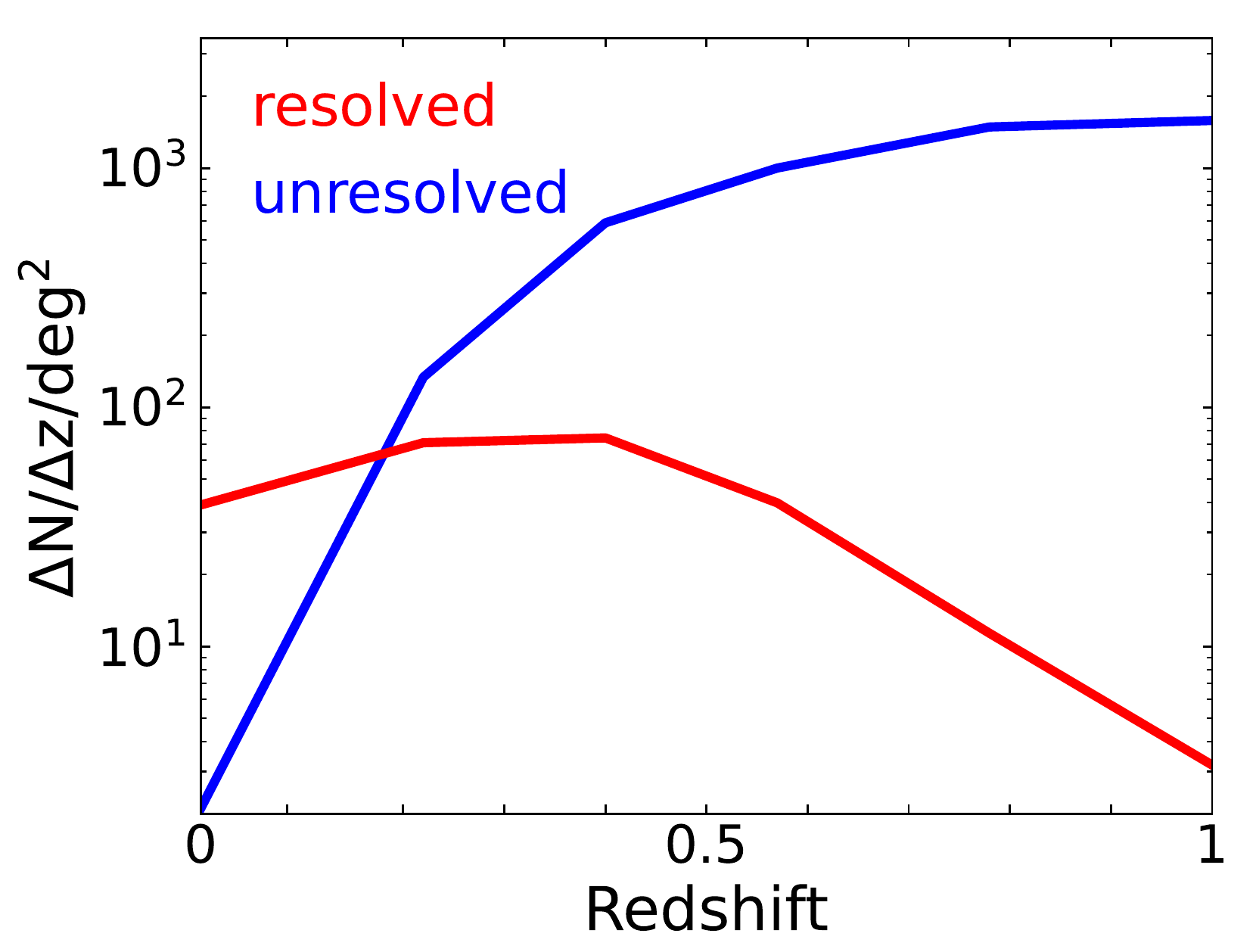}\\
 \caption{The redshift distribution per redshift bin ($\Delta z=0.2$) per square degree of the clusters with $\rm S/N>10$ inside $R_{500}$, averaged through the entire mock sky up to $z=1$. The red curve is the number of extended sources (the angular diameter of $R_{500}$ greater than 3~arcmin), and the blue curve is the number of point-like sources unresolvable by \emph{HUBS} (angular diameter of $R_{500}$ smaller than 3~arcmin).
 }\label{fig:dndzhubs}
\end{figure}

In Fig. \ref{fig:dndzhubs}, we show the redshift distribution per FoV of the clusters with $\rm S/N>10$ inside $R_{500}$, averaged through the entire mock sky up to $z=1$. The red curve is the number of extended sources, and the blue curve is the number of point-like sources unresolvable by \emph{HUBS}.
We can see that the number density of resolved clusters at $z=0$ is about $40~\deg^{-2}$ in a redshift bin of $\Delta z=0.2$. The values peaks at $z\sim0.4$ with almost $80~\deg^{-2}$ per $\Delta z$ and drops rapidly at $z>0.5$ for the decrease in the angular size. Thus, the survey of hot baryons in resolved clusters by \emph{HUBS} should be effective below redshift 0.5 because of the angular size of the clusters in soft X-ray band at different redshift.

\begin{figure*}
\centering
 \includegraphics[angle=0,scale=0.35]{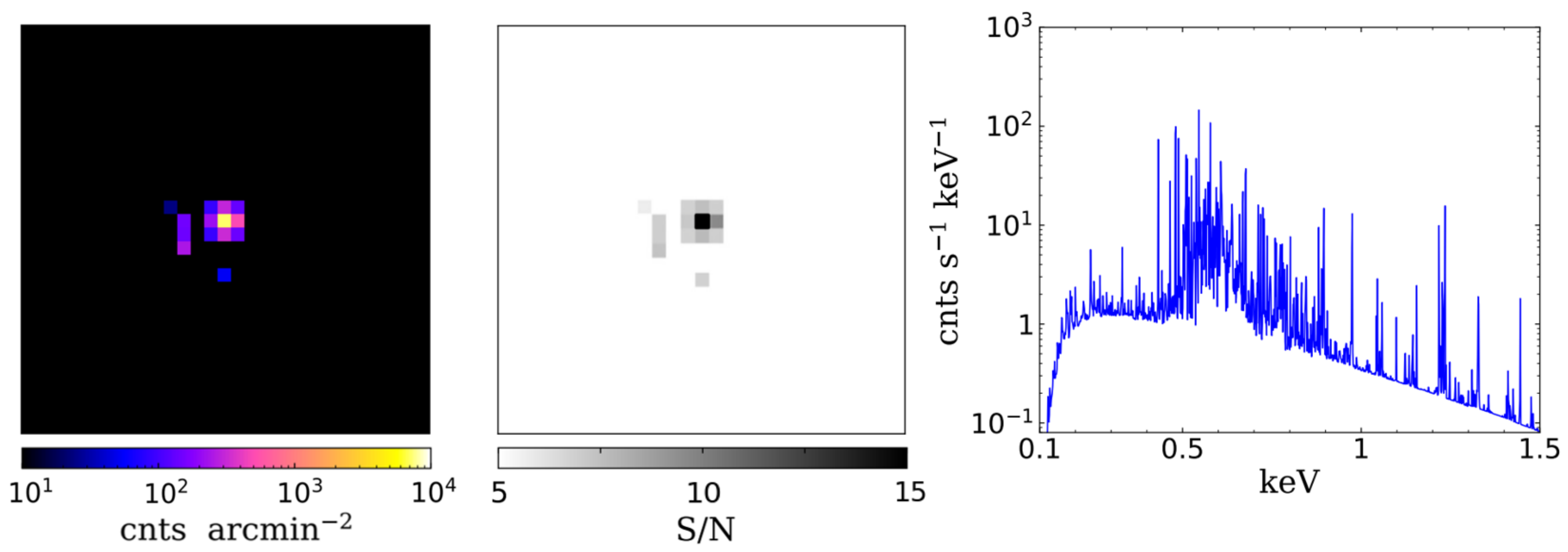}\\
 \caption{The mock observation of \emph{HUBS} for a model cluster at $z\sim0.5$, which is same as the cluster in the middle column of Fig. \ref{fig:imgspec}. The left panel is the photon-count image without the photons from XRB, and the middle panel is the signal-to-noise ratio map. In both panels, the exposure time is $10^6$~s, and the FoV is zoomed-in to $0.5^{\circ}\times0.5^{\circ}$. The right panel is the mock spectrum of this cluster.
 }\label{fig:hubshighz}
\end{figure*}

To test the redshift limit of resolved sources for \emph{HUBS}, we select a massive bright cluster with $L_X>10^{45}~\rm erg~s^{-1}$ at $z\sim0.5$ and show its mock observations in Fig. \ref{fig:hubshighz}. Comparing the emissivity map of the cluster in Fig. \ref{fig:imgspec} and the photon-count image in the left panel of Fig. \ref{fig:hubshighz}, we can see that the selected cluster is close to the angular resolution limit of \emph{HUBS}. In the middle panel of Fig. \ref{fig:hubshighz}, the $\rm S/N$ map indicates that the hot gas in the cluster at redshift around 0.5 can still be detected with an exposure time $10^6$~s, which is consistent with the results in \citet{Zhang2022} that \emph{HUBS} can detect groups and clusters beyond $z\sim0.3$. In addition, the mock spectrum in the right panel indicates that it is also possible for \emph{HUBS} to resolve the strong emission lines in the bright cluster at $z\sim0.5$, and the flux rate of the XRB photons is below $10^{-2}$~cnts~s$^{-1}$~keV$^{-1}$ (not plotted in Fig. \ref{fig:hubshighz}). 

\begin{figure}
\centering
 \includegraphics[angle=0,scale=0.45]{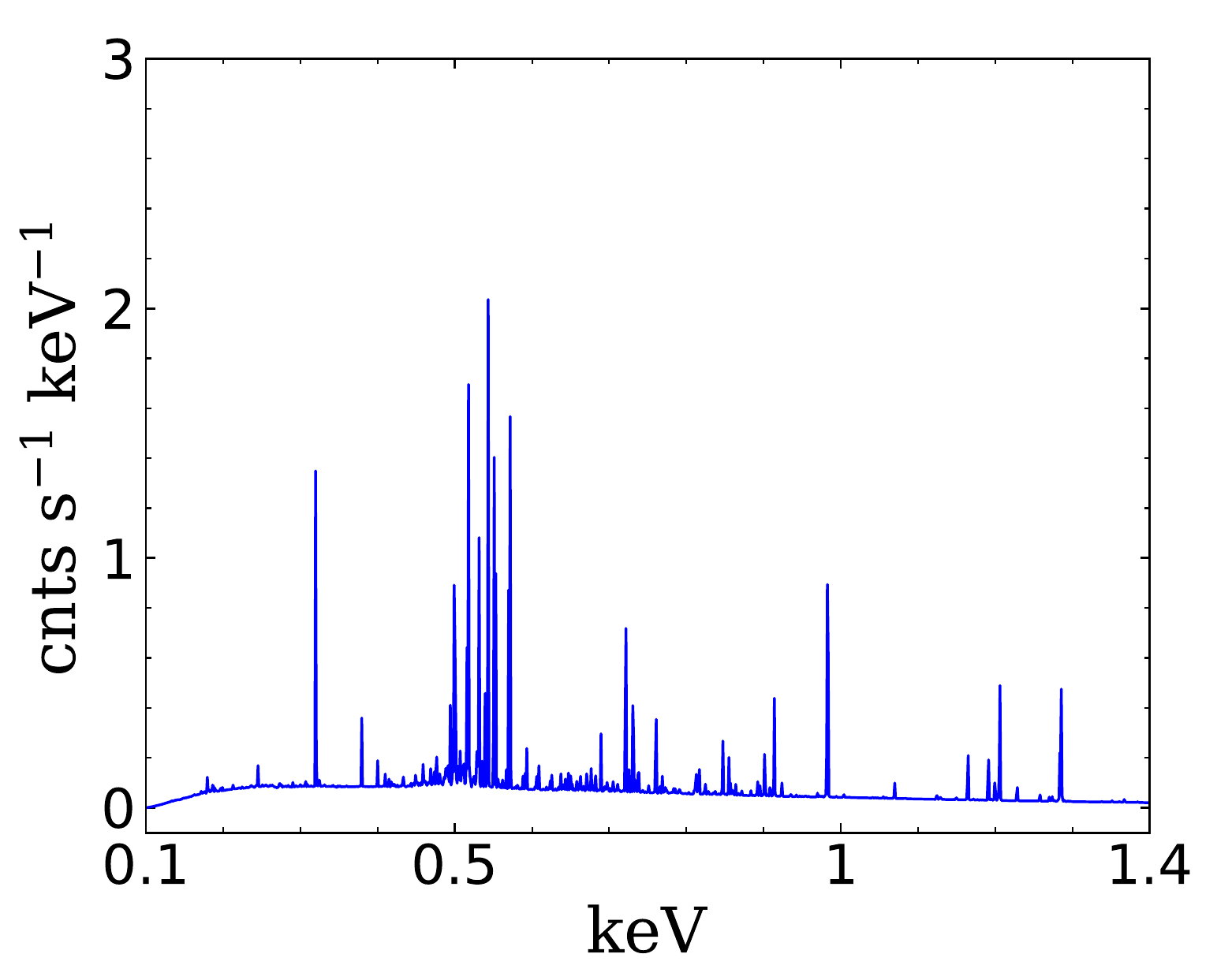}\\
 \caption{The mock spectrum of a cluster with $\M200\sim 3\times10^{13}\ms$ at $z=1.04$. The Y-axis has a linear scale.
 }\label{fig:hubsspez1}
\end{figure}

On the other hand, Fig. \ref{fig:dndzhubs} shows that the number of unresolved sources is around zero at $z=0$ and increases with redshift. It exceeds the number of resolved clusters at $z>0.3$ and reaches around $1000~\deg^{-2}$ per $\Delta z$ at $z>0.8$. These unresolved sources are the clusters with angular size below the angular resolution limit. Because of the large number of these point-like sources, the hot gas in these clusters contributes a significant fraction of baryons at $z\gtrsim0.3$. It is interesting to test the mock observations of the unresolved clusters. 

We select an unresolved cluster at $z\sim 1$ with high signal-to-noise ratio. The halo mass $\M200$ of the cluster is around $3\times10^{13}\ms$, and its angular diameter of $R_{500}$ is around $1.2$~arcmin. After a $10^6$~s of observation by \emph{HUBS}, about $8\times10^{4}$ photons can be detected in 0.1-2~keV band. The mock spectrum of this cluster is in Fig. \ref{fig:hubsspez1}. Although the point-like source at $z\sim1$ is below the angular resolution limit, \emph{HUBS} still has the ability to detect strong emission lines from such kind of source, like the \ion{O}{viii} and \ion{Ne}{x} lines around 0.3 and 0.5~keV in the observed-frame. It should be valuable to observe some sky areas with long exposure time to get the signals from point-like sources of clusters at $z>0.5$, which helps to study the properties and redshift evolution of hot baryons in the early universe. 

In summary, by taking the advantage of the large simulation box in SAMs, the mock observation of \emph{HUBS} will help on the target selection and observation strategies for future survey. Considering the angular size of the clusters, the survey of hot baryons in resolved clusters by \emph{HUBS} is effective below redshift 0.5. \emph{HUBS} has the ability to detect the emission lines of hot gas in clusters at $z>0.5$ and the observation of point-like sources with long exposure time can be used to study the hot baryons in the early universe.

\section{Summary}

In this paper, we create mock X-ray observations of hot gas in galaxy clusters based on the model outputs of a new extension of L-Galaxies SAMs in our recent work in Paper I. Firstly, we use the coordinates and velocities in the model outputs to build some mock light cones up to nearby and deep redshifts. In each light cone, we use the bolometric X-ray flux, gas temperature and gas metallicity to generate mock X-ray spectra for galaxy clusters with SOXS package, and then derive the mock X-ray images of each cluster based on the spectra and the projected X-ray luminosity profiles. Using the mock data, we simulate the X-ray spectra for \emph{ROSAT} all-sky survey, and compare them with the observational results. Then, we consider the design parameters of \emph{HUBS} mission and simulate the observation of hot gas for \emph{HUBS} to evaluate the results for future survey of hot baryons, which is an important application of our mock work.

The main conclusions of this paper are:

\noindent (i) Our mock X-ray observations of hot gas can approximately match the results from X-ray telescopes.

\noindent (ii) Due to the angular size of the clusters, the survey of hot baryons in resolved clusters by \emph{HUBS} is effective below redshift 0.5. \emph{HUBS} has the ability to detect the emission lines of hot gas in clusters at $z>0.5$, and the observation of point-like sources with long exposure time can be used to study the hot baryons in the early universe.

\noindent (iii)  The mock X-ray observations provide the opportunity to make target selection and optimize the observation strategies for forthcoming X-ray facilities by taking the advantage of the large simulation box and flexibility in SAMs.

This paper demonstrates a few applications to use our mock data of hot gas, and many upcoming studies can be carried out in the future. One possible work is the end to end simulation of all-sky hot gas survey of \emph{HUBS} and \emph{eROSITA} considering various systematic and instrumental effects, such as the background sources of AGN, point spread function, redistribution matrix file etc, which provides the sources selection and detection functions at different redshift. Another possible work is to create mock catalogues with SAMs outputs based on ELUCID \citep{Wang2016}, a constrained N-body simulation capable of reproducing the spatial distribution of nearby galaxies and clusters in the real universe, and to simulate the X-ray observations of clusters in given positions of the real sky.

In future SAMs work, it is also necessary to improve the physical prescriptions of the hot gas and X-ray emission, including the X-ray emission from AGN to improve the scaling relations in bright clusters, the cooling and feedback processes in inner haloes to improve the density profiles in the core regions of clusters, and also the distribution of hot baryons beyond the halo viral radius $R_{200}$, which is proposed to be important for the missing baryons in hydrodynamic simulations (e.g \citealt{Martizzi2019}; \citealt{Ayromlou2022}).

\noindent
\\
{\bf Acknowledgments}
The authors thank the anonymous referee for the helpful suggestions. We acknowledge the support from the National SKA Program of China No. 2020SKA0110102, the fund for key programs of Shanghai Astronomical Observatory E195121009, and Shanghai Committee of Science and Technology grant No.19ZR1466700. FY is supported in part by the Natural Science Foundation of China (grants 12133008, 12192220, and 12192223). We thank Dr. Zheng Yunliang in Shanghai Jiao Tong University for his help with the eFEDS data. We thank Prof. Cui Wei in Tsinghua University for his suggestion on carrying out the work in this paper.

\end{document}